# The Methods to Improve Quality of Service by Accounting Secure Parameters

Tamara Radivilova [0000-0001-5975-0269], Lyudmyla Kirichenko [0000-0002-2780-7993], Dmytro Ageiev [0000-0002-2686-3854], Vitalii Bulakh

Kharkiv National University of Radio Electronics, Kharkiv, ave.Nauki, 14, Ukraine
tamara.radivilova@gmail.com

**Abstract.** A solution to the problem of ensuring quality of service, providing a greater number of services with higher efficiency taking into account network security is proposed. In this paper, experiments were conducted to analyze the effect of self-similarity and attacks on the quality of service parameters. Method of buffering and control of channel capacity and calculating of routing cost method in the network, which take into account the parameters of traffic multifractality and the probability of detecting attacks in telecommunications networks were proposed. The both proposed methods accounting the given restrictions on the delay time and the number of lost packets for every type quality of service traffic. During simulation the parameters of transmited traffic (self-similarity, intensity) and the parameters of network (current channel load, node buffer size) were changed and the maximum allowable load of network was determined. The results of analysis show that occurrence of overload when transmitting traffic over a switched channel associated with multifractal traffic characteristics and presence of attack. It was shown that proposed methods can reduce the lost data and improve the efficiency of network resources.

## 1 Introduction

A computer network is a complex and expensive system that solves critical tasks and serves many users. Characteristics of service quality reflect critical network properties: performance, reliability and security [1]. Qualities of service (QoS) methods ensure the stable operation of modern serveces: IP-telephony, video and radio broadcasting, interactive distance learning, etc. QoS methods are aimed at improving the performance characteristics and network reliability and reduce variations in delays and packet loss during periods of overload network [2-3].
Quality of service characteristics reflect the negative impact of the queues mechanism on traffic transmission. This effect can be expressed in a temporary decrease in the rate of traffic transmission, in the delivery of packets with variable delays and in the loss of packets due to a buffer nodes overload. QoS methods are aimed at compensat-ing the negative effects of temporary overloads ocurring in packet-switched networks. These methods use various queue management, reservation and feedback algorithms to reduce the negative impact of queues to acceptable to user's level [3-6].

Experimental and numerical studies conducted in recent decades indicate that many multiservice networks traffic has self-similar properties [2, 7-10]. Self-similar traffic causes significant delays and packet losses, even if the total intensity of all flows is far from the maximum allowable values [7, 10, 11].

A big problem for service providers is to ensure QoS in terms of self-similar traffic, avalanche traffic of intruders, the sources of which are various temporal nodes. This type of behavior is associated with threats such as distributed denial of service (DDoS) attacks, Internet worms, phishing, viruses, email spam and others [13, 14]. The amount of traffic that is generated due to infection, and subsequent traffic bursts can disrupt the normal work of network and create an additional risk to network devices (routers, switches). Security is becoming a critical characteristic of all services and plays a crucial role in the profitability of service providers [15, 16].

When the transfer rate reaches several gigabits, to suppress emerging threats providers must supply protection that ensure reliability and does not affect network performance as a whole [16-18]. The task of functioning to ensure network security determines how a provider can effectively offer a larger volume of services with higher efficiency with a greater degree of manageability [19-21].

The aim of this work is to develop a method of buffering and control channel capacity and routing method in the network, which are based on the multifractal properties of traffic and parameters of network security.

## 2 Self-similar and multifractal traffic's properties

The self-similarity of random processes lies in preservation of probabilistic characteristics when changing the time scale. A stochastic process $X(t)$ is self-similar with a parameter $H$ if the process $a^{-H}X(at)$ is described by the same laws of finite-dimensional distributions as $X(t)$ $Law\{a^{-H}X(at)\} = Law\{X(t)\}, \forall a > 0$, where the parameter $H$, $0 < H < 1$ represents the degree of self-similarity of process and calls the Hurst parameter. The parameter $H > 0.5$ characterizes the measure of the long-term dependence of the stochastic process. The initial moments of self-similar random process can be expressed as $M\left[|X(t)|^q\right] = C(q) \cdot t^{qH}$, where the value $C(q) = M\left[|X(1)|^q\right]$. [5].

Multifractal traffic can be defined as the expansion of self-similar traffic by taking into account the scalable properties of statistical characteristics of second and higher orders. For the moments of multifractal processes, the relation $M\left[|X(t)|^q\right] = c(q) \cdot t^{qh(q)}$ is executed, where $c(q)$ is some deterministic function;

$h(q)$ is the generalized Hurst index which is a nonlinear function in the general case. Value $h(q)$ at $q = 2$ coincides with the value of the degree of self-similarity $H$.

Multifractal traffic has a special structure that persists on many scales: there is always some number of very large bursts with a relatively small average traffic level. As a characteristic of multifractal traffic, it was proposed to use the following parameters: traffic intensity $\lambda$, Hurst parameter $H$ and the range of generalized Hurst index $\triangle h = h(q_{\min}) - h(q_{\max})$. For monofractal processes, the generalized Hurst index does not depend on the parameter $q$ and is a straight line: $h(q) = H$, $\triangle h = 0$. The more process heterogeneity, i.e. the higher emissions are in traffic, leads to the greater range $\triangle h$. The degree of bursts corresponds to the gravity tails distribution. The coefficient of variation $\sigma_{\text{var}}$ can be considered as the simplest quantitative characteristic of the tail distribution $\sigma_{\text{var}}(T) = \sigma(T)/M(T)$, where $T$ is a random variable the values of which are the number of events in a given time interval. [5].

A node can be represented as a queuing system, at each unit of time it receives input data that arrives during this system working period, processes and sends them. The network traffic processing system in this paper is a node with adjustable performance or the data processing speed (the number of packets per unit of time that it can process) and the buffer memory (buffer) with the specified volume (the amount of data it can hold) in which the node puts traffic that it did not manage to process considered step (unit of time) of this work [22-24].

Intrusion detection systems are used to detect network attacks. Their performance and efficiency are evaluated using the parameters of cost, resource utilization, detection rate. Moreover, if it is possible to classify attacks and normal traffic (events), then these are observable parameters. Depending on the nature of the attack and the probability of its detection, four possible outcomes are used [13, 15, 16]:

- True positive (TP): process that is actually an attack and are successfully classified and called attack.
- False Positive (FP): A normal and legitimate process is classified as an attack.
- True Negative (TN): a process that is actually normal and legitimate, and successfully marked and detected as normal.
- False negative (FN): attack is incorrectly classified as normal or legitimate action.

The main difficulty lies in the fact that the number of false positive detections is very large. It is clear that a high FP value will lead to less effective detection, and a high FN value will make the system vulnerable to intruders.

## 3    Method of buffering and control of channel capacity

The simulation is a main tool for research networks with self-similar flows. The simulations of channel load and the queues formation in the buffer for the realization of fractal traffic were studied in [12, 15, 20]. The simulation results allow to calculate the dependence of the values $Net_i^{k_i}(T)$ network bandwidth of $k$ channel of $i$-th node and parameters of incoming data flows $\{\lambda, H, \sigma_{\text{var}}, P_{\text{sec}}\}$ on the size of the buffer memory $Q_w^{new} = f(Net_i^{k_i}, \lambda, H, \sigma_{var}, P_{sec})$, where $P_{sec} = (P_{TP}, P_{FP}, P_{TN}, P_{FN})$ is the probability of detection attacks on system resources (i.e. buffer overload, DDoS attacks). The $P_{sec} = (P_{TP}, P_{FP}, P_{TN}, P_{FN})$ calculates by using the Machine Learning algorithm that shown in [16, 21, 25-27]. Each input flow is sent to a queue $Q_w$ of limited size. Queuing time is dependent on $qs$. The normal traffic transmited through the communication node is provided by the calculated values of the buffer size $Q_w^{new}$ (the loss does not exceed the specified percentage). Similarly, a functional dependence of a specified buffer size $Q_w$ and traffic parameters on a channel capacity $Net_i^{newk_i} = \varphi(Q_w, \lambda, H, \sigma_{var}, P_{sec})$ were obtained. It is possible to avoid network overload by managing of buffers size and / or data flows based on the calculation of the value of the maximum permissible load in accordance with the obtained results. To determine maximum allowable load of the channel for the given size of channel capacity and the buffer memory are used the functional dependencies $Net_i^{newk_i} = \varphi(Q_w, \lambda, H, \sigma_{var}, P_{sec})$ and $Q_w^{new} = f(Net_i^{k_i}, \lambda, H, \sigma_{var}, P_{sec})$. By predicting the start of overload, the required size of the channel capacity $Net_i^{newk_i}$ and / or buffer memory $Q_w^{new}$ can be allocated. If the calculated buffer size $Q_w^{new}$ is larger than the existing $Q_w$ and the probability of detection attacks on system resources $P_{sec} = (P_{TP}, P_{FP}, P_{TN}, P_{FN})$ is high the system allocated the required size of buffer memory. But if probability of detection attacks on system resources $P_{sec} = (P_{TP}, P_{FP}, P_{TN}, P_{FN})$ is low the system creates the alert and doesn't allocate the required size of buffer memory. The required size of channel capacity can be determined from the received data. If the requested capacity size $Net_i^{newk_i}$ is larger than available $Net_i^{k_i}$ and the probability of detection attacks on system resources $P_{sec}$ is high, the system provides the requested resource, thus distributing the rest to the second channel capacity. But if $P_{sec}$ is low the system created the alert and doesn't provide the requested resource.

This method can be used in nodes (switch, router and i.e.) for preventing of network overload. It allows to reduce a loss packets and increase channel utilization and network performance.

## 4    Calculating of routing cost method in the network

To provide QoS is required routes selection based on separate safety flows, at the same time, different flows that are sent to one recipient may be directed by various pathes. In addition, the paved pathes can be changed in case of overload. The shortest paths (Low Cost Routing) between the incoming edge router and others are calculated by the routing protocol. Lets consider the calculating of routing cost method based on security parameters and fractal structure of traffic [4, 23, 25, 26].

The communication links between nodes with the maximum capacity $Link_{lk} = \{L_{lk}\}$, $lk = 1, 2, ..$, that are divided into $k \in K$ channels with bandwidth $Net_{lk}^k(t) = \{Net_{lk}^k\}$ at time $t$ [15, 26, 27]. Assume that at each time $t$ traffic of intensity $\lambda_{Net_i}^{qs}(t)$ relating to one of the classes of service $qs$-th with requirements QoS. Each $qs$ corresponds to the maximum percentage of loss $l_{qs}$ and the maximum delay value $\tau_{qs}$. The variable $X_i^{qs}(t) \in X$ is the coefficient of loss $qs$ traffic that transmitted by the path $Net_{lk}^k(t)$ to the $i$-th node at the moment $t$. It is assumed that the probability of package error in the path can be neglected and losses occur only in node because buffer overloads. The coefficient of losses for all nodes in the network:

$$0 \leq X_i^{qs}(t), \sum_{i=1}^{N} X_i^{qs}(t) \leq l_{qs}. \qquad (1)$$

Thus, the restriction (1) show that the total loss for the traffic $\lambda_{Net_i}^{qs}(t)$ routed at time $t$, should not exceed the maximum permissible values for the class of service $l_{qs}$. Loss is defined as the ratio of the discarded data to received data. Value $X_i^{qs}(t) \to \min$ is subject to minimization.

Restrictions imposed by the delay time are similar: $0 \leq T_i^{qs}(t), \sum_{i=1}^{N} T_i^{qs}(t) \leq \tau_{qs}$, where $T_i^{qs}(t)$ is the average waiting time of package $qs$-th class of service in the queue at the $i$-th node [21, 28, 29]. Performing this restriction helps to ensure that the packets delivery does not exceed the maximum permissible values for a given class of service $\tau_{qs}$.

All input traffic is divided at the $qs$ flows so that to ensure the transmission requirements of all classes $QS(t)$ in full. The channels set of traffic QoS $K = K(Net_{lk}^k, P_{lk}, L_k)$, where $P_{lk} = \left\{ p_{lk}^1, ..., p_{lk}^k \right\}$ is allowable set of pathes to the path $L_k$, that is defined for each traffic channel.

The value of routing cost $c_{lk}$ is assigned to the communication link $lk$ and may depend on several parameters, particularly the reliability, speed, and length. The cost of path $p_{lk}^k$ is equal to the sum of communication lines cost: $C_{lk}^k = \sum_{lk \in p_{lk}^k} c_{lk}$. If $Netx_{lk}^k(t)$ is bandwidth that is forwarded to allowable path $p_{lk}^k$ of transmission traffic channel $\lambda_{Net_i}^{qs}(t)$, then following relation holds $\sum_{t \in T; lk=1}^{L_k} Netx_{lk}^k(t) = Net_{lk}^k$, $\forall k \in K$, $\forall lk \in \{1, ..., L_k\}$.

In [10, 13, 15] it was shown that at values $H \geq 0,9$ or at $H > 0,5$ and simultaneously values $\sigma_{var} \geq 3$ (which roughly corresponds to the values $\Delta h > 1$) the amount of loss is greater than 5-10%. When passing traffic with strong fractal properties it needs to timely increase bandwidth of communication lines. To reflect the changes in the multifractal properties of flows, cost of paths $C_{lk}^k$ are updated in regular intervals and recalculated by the formula

$$Cnew_{lk}^k = \begin{cases} C_{lk}^k, & H \leq 0,5; \quad P_{sec} < 0.6 \\ C_{lk}^k + (H - 0.5)C_0, & 0.5 < H < 0.9, \sigma_{var} \leq 1, P_{sec} > 0.6; \\ C_{lk}^k + (H - 0.5)(\sigma_{var} - 1)C_0, & 0.5 < H < 0.9, 1 < \sigma_{var} < 3, P_{sec} > 0.6; \\ C_{lk}^k + C_0, & H \geq 0.9 \text{ or } H > 0.5, \sigma_{var} \geq 3, P_{sec} > 0.6. \end{cases}$$

where $C_{lk}^k = \sum_{lk \in p_{lk}^k} c_{lk}$ is determined in accordance with the objective function, value $C_0$ is selected by the network administrator considering network topology and probability of detection attacks on system resources $P_{sec} = (P_{TP}, P_{FP}, P_{TN}, P_{FN})$. The routing algorithm is not changed (path cost $Cnew_{lk}^k = C_{lk}^k$) if the traffic has independent values ($H = 0.5$) or has antipersistent properties ($H < 0.5$) and $P_{sec} < 0.6$. If the value $0.5 < H < 0.9$ and the dispersion of data is small ($\sigma_{var} \leq 1$) the value $C_{lk}^k$ increases in proportion to the Hurst exponent value. If the Hurst exponent value $0.5 < H < 0.9$, $P_{sec} > 0.6$ and dispersion is large ($1 < \sigma_{var} < 3$) the value $C_{lk}^k$ increases in proportion to three characteristics. The cost with a maximum value $C_{lk}^k + C_0$ is obtained at $H \geq 0.9$, $P_{sec} > 0.6$ or persistent traffic ($H > 0.5$) with a coefficient of variation $\sigma_{var} \geq 3$. After recalculating the value of all paths, the announcement of the state of paths are sent between routers.

This method can be used to increase utilization channel by rerouting most important data flows to alternative low load channels.

## 5 Simulation of the Developed Methods

To carry out simulation work of the proposed methods program modules were developed by using Python. The input system receives the generated information flows having predetermined fractal properties with parameters similar to the real traffic [22, 25-27, 29]. This traffic was sent from the sender to receiver through 14 nodes: routers, nodes and firewall by using various parthes. By using model realizations traffic with predetermined properties in experiments a various network parameters can be determined in a variety of loading and operation modes. The network load was changed from 20% to 90%. All traffic parameters were changed and some of them have attacks.

The dynamic control of channels capacity and sizes of nodes buffer memory were performed during simulation with method of buffering and control channel capacity (MBCCC).

In the dynamic channel capacity control mode, a forecast of flows requirements in the necessary channel capacity $Net_i^{newk_i}$ were made in the subsequent time interval t+Δ (ms). Based on these data the necessary resource of capacity $Net_i^{newk_i}$ was allocated for self-similar traffic critical to the time of transmission and the remaining traffic by the remaining resource for a time Δ.

Similarly, in the buffer size control mode of nodes the buffer memory requirements of the node $Q_w^{new}$ were calculated for a time t+Δ. Based on this data required resources of buffer memory $Q_w^{new}$ (MB) node were allocated for a while.

During the simulation with Routing method (RM) calculation of routing costs $Cnew_{lk}^k$ of fractal traffic in channels at the time t were calculated taking into account the required throughput $Net_{lk}^k(t)$, using the minimum cost criterion and probability of detection attacks $P_{sec}$, with limiting the QoS for the uniform use of traffic channels of different QoS classes in the case of multifractal flows. The fractal traffic of high $qs$ class of service was routed along most free, secure, lowest cost paths taking into account traffic properties. It is possible to multiplex low-priority traffic into separate flows so that its life time does not expire.

The effectiveness of proposed methods we can evaluate by analysis of the simulation results: changing a buffer memory size and the dynamic distribution of a channel capacity.

During the experiment, the amount of channels utilization, value of lost data, value of jitter, probability of detection attacks on system resources $P_{sec}$ were measured. The QoS parameters that were obtained during the experiments by using proposed methods are shown on Figure 1 and in Table 1. The channel utilizations are shown in Figure 1 (a) and lost packets are shown on Figure 1 (b).

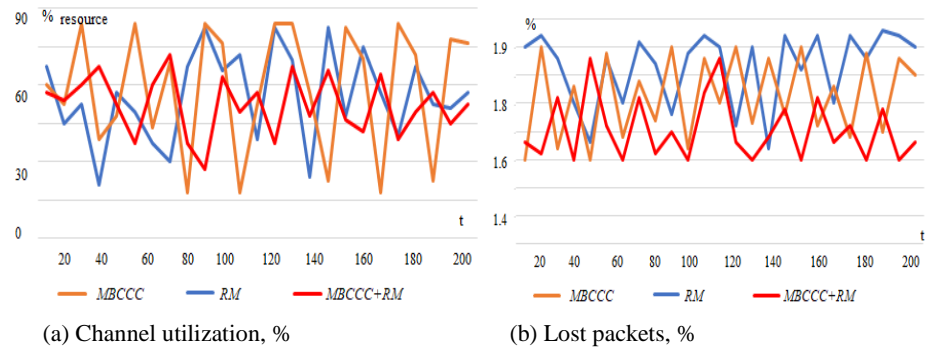

(a) Channel utilization, %  (b) Lost packets, %

**Fig.1.** QoS parameters by using the proposed methods.

**Table 1.** The parameters of quality of service.

| Methods | Channel utilization, % | Value of lost data, % | Jitter, ms | $P_{sec}$ (TP), % | $P_{sec}$ (FP), % |
|---|---|---|---|---|---|
| Method of buffering and control of channel capacity | 0.69 | 1.88 | 54 | 0.62 | 0.31 |
| Routing method in the network | 0.64 | 1.9 | 38 | 0.65 | 0.33 |
| Using both methods | 0.52 | 1.73 | 33 | 0.71 | 0.28 |

With the same volume of transmitted information in network the lost of self-similar traffic during transmission is noticeably lower with simultaneous use of proposed methods. It is similarly for parameters of channel utilization and jitter. The probability of attack detection is improved when both proposed methods are used.

The results of the experiments were realized and they brought the following effect: increasing the utilization of data transmission channels, by redirecting the most critical information flows to less loaded alternative channels, more efficient usage of system resources, improving the quality of service, and reduction of data loss.

## 6       Discussion and Conclusion

In this paper by collecting and analyzing traffic in real-time the network QoS parameters were improve. Methods for ensuring QoS are proposed. They are taking into account the parameters of the probability of attacks detection in telecommunications networks during the transmission of fractal traffic over the network. The occurrence of congestion during the transmit of traffic through a switched communication channel is associated with the multifractal characteristics of traffic. Depending on the multifractality parameters, the current channel load, the node buffer size it is possible to determine the maximum permissible load of this network.

A method for predicting network congestion based on the calculation of the degree of communication channel load when traffic monitoring considering the probability of detection attacks. A method for estimating the cost of routing is described and based on accounting the fractal properties of network traffic, security parameters and specified restrictions on delay time and the number of lost packets.

The proposed methods can reduce network data loss and increase the efficiency of network resources, provide higher performance and throughput, reduce costs by redirecting the most critical information flows to less loaded alternative channels, and reduce response time and the amount of lost data. In future work it is planned to investigate the influence and dependence of various types and levels of attacks on the network QoS parameters.

**References.**

1. Yu-Xing, Y.: Research for Network Security and Reliability and Performance Assess. In: Proceedings of the Fourth International Conference on Intelligent Systems Design

and Engineering Applications, Zhangjiajie, pp. 444-447 (2013). doi: 10.1109/ISDEA.2013.506
2. Daradkeh, Y. I., Kirichenko, L., Radivilova, T.: Development of QoS Methods in the Information Networks with Fractal Traffic. Intl. Journal of Electronics and Telecommunications 64 (1), 27-32 (2018). doi: 10.24425/118142
3. Acharya, H. S., Dutta, S. R., Bhoi, R.: The Impact of self-similarity Network traffic on quality of services (QoS) of Telecommunication Network. International Journal of IT Engineering and Applied Sciences Research (IJIEASR) 2, 54-60 (2013)
4. Czarkowski, M., Kaczmarek, S., Wolff, M.: Influence of Self-Similar Traffic Type on Performance of QoS Routing Algorithms. INTL Journal of electronics and telecommunications 62(1), 81-87 (2016). doi: 10.1515/eletel-2016-0011
5. Ivanisenko, I., Kirichenko, L., Radivilova, T.: Investigation of self-similar properties of additive data traffic. 2015 Xth International Scientific and Technical Conference "Computer Sciences and Information Technologies" (CSIT) Proceedings, pp. 169-171. IEEE (2015). doi: 10.1109/STC-CSIT.2015.7325459
6. Pramanik, S., Datta, R., Chatterjee, P.: Self-similarity of data traffic in a Delay Tolerant Network. In: 2017 Wireless Days, pp.39–42 (2017). doi: 10.1109/WD.2017.7918112
7. Bhat, U.N.: An Introduction to Queueing Theory: Modeling and Analysis in Applications. Birkhauser, Springer (2015)
8. Hu, Z., Mukhin, V., Kornaga, Y., Lavrenko, Y., Herasymenko, O.: Distributed Computer System Resources Control Mechanism Based on Network – Centric Approach. I.J. Intelligent Systems and Applications 7, 41-51 (2017). doi: 10.5815/ijisa.2017.07.05
9. Ivanisenko, I., Radivilova, T.: The multifractal load balancing method. Second International Scientific-Practical Conference Problems of Infocommunications Science and Technology (PIC S&T) Proceedings, pp. 122-123. IEEE (2015). doi: 10.1109/INFOCOMMST.2015.7357289
10. Aggarwal, A., Verma, R., Singh, A.: An Efficient Approach for Resource Allocations using Hybrid Scheduling and Optimization in Distributed System. I. J. Education and Management Engineering 3, 33-42 (2018). doi: 10.5815/ijeme.2018.03.04
11. Radivilova, T., Kirichenko, L., Ivanisenko, I.: Calculation of distributed system imbalance in condition of multifractal load. Third International Scientific-Practical Conference Problems of Infocommunications Science and Technology (PIC S&T) Proceedings, pp. 156-158. IEEE (2016). doi: 10.1109/INFOCOMMST.2016.7905366
12. Kirichenko, L., Ivanisenko, I., Radivilova, T.: Dynamic load balancing algorithm of distributed systems. In: Proceedings of the 13th International Conference on Modern Problems of Radio Engineering, Telecommunications and Computer Science (TCSET) Proceedings, pp. 515-518. IEEE (2016). doi: 10.1109/TCSET.2016.7452102
13. Gupta, N., Srivastava, K., Sharma, A.: Reducing False Positive in Intrusion Detection System: A Survey. International Journal of Computer Science and Information Technologies 7 (3), 1600-1603 (2016)
14. Deka, R., Bhattacharyya, D.: Self-similarity based DDoS attack detection using Hurst parameter. Security and Communication Networks 9(17), 4468-4481 (2016)
15. Ageyev, D., Kirichenko, L., Radivilova, T., Tawalbeh, M., Baranovskyi, O.: Method of self-similar load balancing in network intrusion detection system. In: 28th International Conference Radioelektronika (RADIOELEKTRONIKA) Proceedings, pp. 1-4. IEEE (2018). doi: 10.1109/RADIOELEK.2018.8376406

16. Bulakh, V., Kirichenko, L., Radivilova, T., Ageiev, D.: Intrusion Detection of Traffic Realizations Based on Maching Learning using Fractal Properties. In: 2018 International Conference on Information and Telecommunication Technologies and Radio Electronics (UkrMiCo) Proceedings, pp.1-4. IEEE (2018)
17. Sharma, R., Chaurasia, S.: An Integrated Perceptron Kernel Classifier for Intrusion Detection System. I. J. Computer Network and Information Security 12, 11-20 (2018). doi: 10.5815/ijcnis.2018.12.02
18. Chahal, J.K., Kaur, A.: A Hybrid Approach based on Classification and Clustering for Intrusion Detection System. I.J. Mathematical Sciences and Computing 4, 34-40 (2016). doi: 10.5815/ijmsc.2016.04.04
19. Nasr, A.A., Ezz, M.M., Abdulmaged, M.Z.: An Intrusion Detection and Prevention System based on Automatic Learning of Traffic Anomalies. I. J. Computer Network and Information Security 1, 53-60 (2016). doi: 10.5815/ijcnis.2016.01.07
20. Kirichenko, L., Radivilova, T.: Analyzes of the distributed system load with multifractal input data flows. In: 14th International Conference The Experience of Designing and Application of CAD Systems in Microelectronics (CADSM) Proceedings, pp. 260-264. IEEE (2017). doi: 10.1109/CADSM.2017.7916130
21. Popa, S. M., Manea, G. M.: Using Traffic Self-Similarity for Network Anomalies Detection. In: 20th International Conference on Control Systems and Computer Science Proceedings, pp. 639-644. IEEE (2015)
22. Betker, A., Gamrath, I., Kosiankowski, D., Lange, C., Lehmann, H., Pfeuffer, F., Simon, F., Werner, A.: Comprehensive topology and traffic model of a nationwide telecommunication network. IEEE/OSA Journal of Optical Communications and Networking 6(11), 1038 – 1047 (2014). doi: 10.1364/JOCN.6.001038
23. Han, D., Chung, J.-M.: Self-Similar Traffic End-to-End Delay Minimization Multipath Routing Algorithm. IEEE Communications Letters 18(12), 2121-2124 (2014). doi: 10.1109/LCOMM.2014.2362747
24. Lemeshko, A. V., Evseeva, O. Y., Garkusha, S. V.: Research on tensor model of multi-path routing in telecommunication network with support of service quality by greate number of indices. Telecommunications and Radio Engineering, 73(15), 1339-1360 (2014). https://doi.org/10.1615/TelecomRadEng.v73.i15.30
25. Naik, L., Khan, R.U., Mishra, R.B.: MANETs: QoS and Investigations on Optimized Link State Routing Protocol. I. J Computer Network and Information Security 10, 26-37 (2018). doi: 10.5815/ijcnis.2018.10.04
26. Lemeshko, O., Yeremenko, O.: Enhanced method of fast re-routing with load balancing in software-defined networks. Journal of Electrical Engineering 68(6), 444-454 (2017). doi: 10.1515/jee-2017-0079
27. Yeremenko, O., Lemeshko, O., Persikov, A.: Secure Routing in Reliable Networks: Pro-active and Reactive Approach. Advances in Intelligent Systems and Computing II, CSIT 2017, Advances in Intelligent Systems and Computing 689, 631-655 (2018). doi: 10.1007/978-3-319-70581-1_44
28. Fan, C., Zhang, T., Zeng, Z., Chen, Y.: Optimal Base Station Density in Cellular Networks with Self-Similar Traffic Characteristics. In: 2017 Wireless Communications and Networking Conference (WCNC) Proceedings pp. 1-6. (2017). doi: 10.1109/WCNC.2017.7925881
29. Bulakh, V., Kirichenko, L., Radivilova, T.: Time Series Classification Based on Fractal Properties. In: Second International Conference on Data Stream Mining & Processing (DSMP) Proceedings, pp. 198-201. IEEE (2018). doi: 10.1109/DSMP.2018.8478532